\documentclass[journal=nalefd,manuscript=article, layout=onecolumn]{achemso}

\usepackage[version=3]{mhchem} 
\usepackage{float}
\usepackage[normalem]{ulem} 
\usepackage[colorinlistoftodos]{todonotes}

\definecolor{cor_color}{RGB}{0,0,0} 
\newcommand{\cor}[1]{{\leavevmode\color{cor_color}{#1}}} 

\usepackage[final]{changes}

\definechangesauthor[name={Thomas Weiss}, color=blue]{TW}
\definechangesauthor[name={Sergei Gladyshev}, color=brown]{SG}

\renewcommand{\citep}[1]{[\!\!\citenum{#1}]}

\author{Sergei Gladyshev}
\affiliation{Institute of Physics, University of Graz, Universitätsplatz 5, and NAWI Graz, 8010 Graz, Austria}

\author{Connor Heimig}
\affiliation{Chair in Hybrid Nanosystems, Nano-Institute Munich, Faculty of Physics, Ludwig-Maximilians-Universität München, 80539 Munich, Germany}

\author{Adrià Canós Valero}
\affiliation{Institute of Physics, University of Graz, Universitätsplatz 5, and NAWI Graz, 8010 Graz, Austria}

\author{Dmytro Gryb}
\affiliation{Chair in Hybrid Nanosystems, Nano-Institute Munich, Faculty of Physics, Ludwig-Maximilians-Universität München, 80539 Munich, Germany}

\author{Tao Jiang}
\affiliation{Chair in Hybrid Nanosystems, Nano-Institute Munich, Faculty of Physics, Ludwig-Maximilians-Universität München, 80539 Munich, Germany}

\author{Angana Bhattacharya}
\affiliation{Chair in Hybrid Nanosystems, Nano-Institute Munich, Faculty of Physics, Ludwig-Maximilians-Universität München, 80539 Munich, Germany}

 \author{Sebastian A. Schulz}
\affiliation{School of Physics and Astronomy, University of St Andrews, North Haugh, St Andrews KY16 9SS, UK}
\alsoaffiliation{Department of Electrical and Computer Engineering, Waterloo Institute of Nanotechnology, University of Waterloo, 200 University Avenue West, Waterloo ON N2L 3G1}

\author{Peter Banzer}
\affiliation{Institute of Physics, University of Graz, Universitätsplatz 5, and NAWI Graz, 8010 Graz, Austria}

\author{Andreas Tittl}
\affiliation{Chair in Hybrid Nanosystems, Nano-Institute Munich, Faculty of Physics, Ludwig-Maximilians-Universität München, 80539 Munich, Germany}
\alsoaffiliation{Institute of Photonics, Hamburg University of Technology, 21073 Hamburg, Germany}

\author{Thomas Weiss}
\affiliation{Institute of Physics, University of Graz, Universitätsplatz 5, and NAWI Graz, 8010 Graz, Austria}
\email{ sergei.gladyshev@uni-graz.at,thomas.weiss@uni-graz.at
}

\title[An \textsf{achemso} demo]
  {Extreme light confinement mediated by the transverse Kerker effect}

\begin{document}

\begin{abstract} 
Dielectric nanoparticles can be engineered to scatter light predominantly in the transverse direction, a phenomenon known as the \replaced[id = TW]{transverse}{Transverse} Kerker effect\deleted[id = TW, comment = {No abbreviations in abstracts!}]{(TKE)}. Although complete \replaced[id = TW]{cancelation}{cancellation} of forward scattering from a single object is forbidden by the optical theorem, we show that a single photonic mode can nonetheless realize an ideal \replaced[id = TW]{transverse Kerker effect}{TKE}. The mode remains dark under normal incidence but evolves into an accidental bound state in the continuum \deleted[id = TW]{(BIC)} when the nanoparticles are arranged in metasurfaces. This enables a new route to polarization-independent quasi-\replaced[id = TW]{bound states in the continuum}{BICs} whose quality factors are tunable without symmetry breaking. We experimentally demonstrate our concept in the visible, achieving the first polarization-independent \replaced[id = TW]{bound state in the continuum}{BIC} without the need for Brillouin-zone folding. Furthermore, we show that our modes maintain large quality factors over a substantially broader region of momentum space than conventional \replaced[id = TW]{bound states in the continuum}{BICs}. Our results establish a platform for realizing ultranarrow resonances free of the constraints \replaced[id = TW]{for designs with standard bound states in the continuum}{of standard BICs designs}.

\end{abstract}

Controlling both the direction and intensity of electromagnetic radiation is instrumental for a wide range of nanophotonic applications, including the design of more efficient metasurfaces~\cite{Khorasaninejad2016Jun,So2023Oct,Azzam2021Mar}, the enhancement of nanolaser performance~\cite{Kodigala2017Jan,Ha2018Nov,Hwang2021Jul}, and the realization of novel photonic devices~\cite{Khanikaev2013Mar,Butow2024Mar}. Such control over radiation patterns enables the creation of highly integrated components characterized \replaced[id=SG]{with improved efficiency.}{by optimal efficiency.} In this context, the Kerker effect~\cite{kerker1983electromagnetic}, manifesting itself as a giant asymmetry between forward-to-backward electromagnetic scattering, is a simple yet powerful mechanism to control the directionality of light scattered by plasmonic and dielectric nanoparticles~\cite{person2013demonstration,nieto2010angle,staude2013tailoring,hancu2014multipolar,fu2013directional,alaee2015magnetoelectric}. While conventional Kerker-type responses control the balance between forward and backward scattering, they cannot \added{simultaneously} provide scattering \replaced[id = TW]{cancelation}{cancellation} in both the forward and backward directions.

Distinct from the \deleted{well-known} Kerker effect, properly-designed dielectric nanoparticles have been shown to strongly suppress both forward and backward scattering while significantly enhancing lateral scattering, offering new avenues for advanced light manipulation~\cite{neugebauer2016polarization,Bag2018Nov,Shamkhi2019Aug,Qin2022May}. This phenomenon is \deleted{now} known as the \replaced[id = TW]{transverse}{Transverse} Kerker effect (TKE).

However, the ideal TKE cannot be realized in localized scatterers under plane wave illumination.  According to the optical theorem~\cite{Bohren1998Apr}, light extinction by an object is proportional to the imaginary part of the light scattering amplitude in the forward direction. As a result, the realization of the TKE is only approximate~\cite{Shamkhi2019Aug} when one considers the scattering of a localized object under plane-wave illumination. 

\begin{figure*}[t]
\begin{center}
\includegraphics[width=1\textwidth]{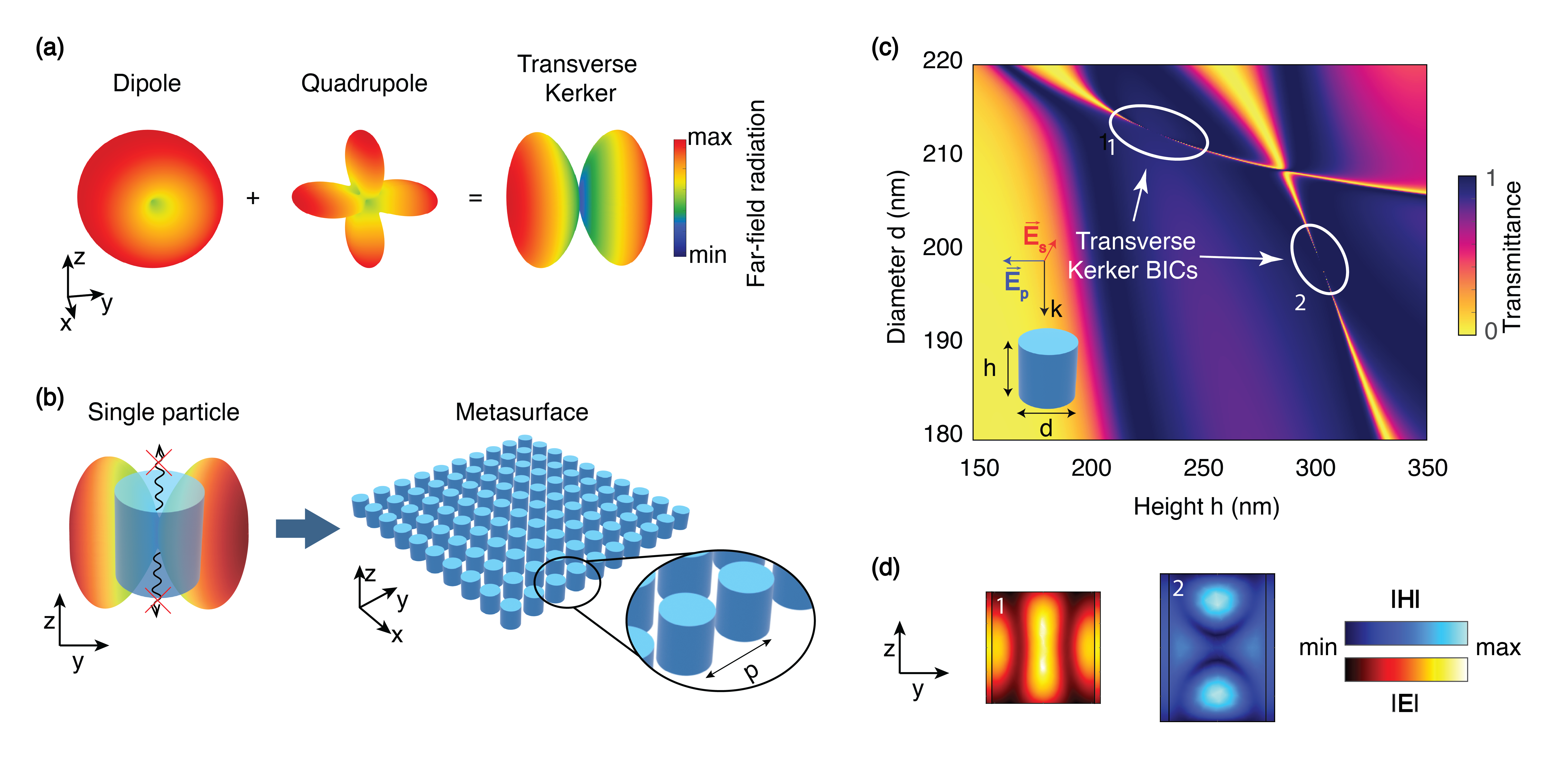}
\caption{\textbf{(a)} Formation mechanism of the \replaced{transverse Kerker effect}{TKE} in terms of multipole moments.   \textbf{(b)} A square lattice of dielectric cylinders. Parameters: refractive index $n=3.85$, period  \replaced[id = SG]{$p$}{$P$} = 235.8 nm. \textbf{(c)} Transmittance map as a function of height \replaced[id = SG]{$h$}{$H$} and diameter  \replaced[id = SG]{$d$}{$D$} of the cylinder at normal incidence of a plane wave, calculated at wavelength $\lambda$ = 645 nm. \textbf{(d)} \replaced{Magnitude of electric and magnetic fields for the transverse-Kerker bound states in the continuum 1 and 2 (see panel c), displayed on planes through the center of the unit cell.}{The distribution of absolute value of electric and magnetic fields for two different  modes supporting TK BIC.} } 
\label{fig:concept}
\end{center}
\end{figure*}

The theoretical framework used to describe the TKE is the multipole expansion of the scattered field~\cite{Shamkhi2019Aug,Bohren1998Apr}. In this approach, the scattered field is decomposed into a series of multipolar contributions represented by vector spherical harmonics \deleted[id = TW]{(VSHs)} of different multipole orders (dipole, quadrupole, and so on) and of electric and magnetic type~\cite{jackson1999}. Crucially, under certain conditions, the destructive interference between electric and magnetic multipoles of consecutive orders can, in principle, lead to transverse asymmetry in the far field scattering pattern, giving rise to the TKE.~\cite{kuznetsov2022special,Qin2022May} To lowest order, the TKE under plane wave illumination originates from the interference of the electric (magnetic) dipole and the magnetic (electric) quadrupole (see Fig.~\ref{fig:concept}\textbf{(a)}). 

\replaced[id=SG]{In contrast to far-field scattering,}{In contrast to the scattering problem,} the multipolar content of the eigenmodes of nanoresonators has been shown to be directly linked to their quality factors (Q~factors).~\cite{sadrieva2019multipolar,Gladyshev2020Aug,chen2019singularities,canos2025exceptional} In extreme cases, the multipoles \replaced[id = TW, comment = {It's better to introduce BIC as abbrevation.}]{result in the formation of a bound state in the continuum (BIC)}{dictate the formation of Bound States in the Continuum (BICs)} in periodic nanophotonic structures.~\cite{sadrieva2019multipolar,chen2019singularities} Recently, BICs have received a great deal of attention since they are characterized by infinite radiative Q~factors, despite lying \replaced[id = TW]{inside}{above} the light cone~\cite{Marinica2008May,Hsu2016Jul,Yu2019Apr,neale2021accidental,Koshelev2023May}. BICs are generally categorized into symmetry-protected and accidental types.~\cite{Hsu2016Jul} In practice, they manifest as quasi-BICs when the resonance linewidth becomes finite~\cite{Koshelev2018Nov,Sadrieva2017Apr} and the modes can be probed from the far-field. They enable ultrasensitive biosensing and molecular handedness detection due to their ultrahigh \replaced[id = TW]{Q~factors}{resonances}~\cite{yesilkoy2019ultrasensitive,chen2020integrated,sharma2025accidental,liu2023phase,shakirova2025molecular}. 
Furthermore, quasi-BICs have been harnessed to realize low-threshold and room-temperature nanolasers~\cite{wu2020room,do2025room,zhou2025ultrahigh}, 
as well as to enhance light emission from quantum materials~\cite{guo2025topologically,lee2025bound}. In addition, their strong field localization has been utilized for advanced photodetection and nonlinear or chiroptical effects in metasurfaces~\cite{wu2025plasmonic,liang2020bound,koshelev2023resonant}.

 In practice, the design of accidental BICs is significantly more challenging than symmetry-protected BICs because of their extreme sensitivity to changes in the geometry of the unit cell or the material parameters. However, such a high sensitivity also renders them highly desirable for sensing, in contrast to symmetry-protected BICs. \replaced[id=SG]{Moreover, the associated quasi-BICs allow for broader tuning of the resonance frequency, linewidth, and coupling strength, without requiring symmetry breaking to become observable.}{Moreover, their associated \replaced[id = TW]{quasi-BICs}{quasi-BIC resonances} present a much larger degree of tunability and do not require symmetry-breaking perturbations to be observed}~\cite{han2024observation,sidorenko2021observation,abujetas2022tailoring,gladyshev2023inverse}. 

In this work, we reveal that the radiation patterns of the eigenmodes of isolated single scatterers can be tuned to \emph{exactly} fulfill the TKE \emph{without breaking the optical theorem}. As a result, such eigenmodes become perfectly decoupled from normally incident light. However, when the nanoparticles supporting them are arranged on a periodic lattice, the resulting collective modes give rise to accidental BICs at normal incidence [see Fig.~\ref{fig:concept}\textbf{(b-c)}]. We term these \textit{transverse-Kerker bound states in the continuum} (TK BICs). While a metasurface with similar behavior was theoretically proposed earlier,~\cite{Allayarov2024Jun} the mechanism giving rise to the ideal TKE in an isolated particle has not been identified yet and hence no guideline for the design of TK BICs was established. Leveraging these insights, we design and fabricate a metasurface of silicon nanodisks that supports a TK BIC in the visible range. 
 
Notably, unlike conventional BICs, the quasi-BICs \replaced[id=SG]{originating}{spawning} from TK BICs are doubly degenerate, and can be excited by a normally incident plane wave of arbitrary polarization. So far, polarization-independent quasi-BICs were only realized in the visible through the mechanism of Brillouin-zone folding,~\cite{yang2025polarization,vaity2022polarization} which requires a complex arrangement of meta-atoms in a single unit cell. This is in strong contrast with the simple architecture demonstrated here, where polarization-independent quasi-BICs are excited in a conventional metasurface of silicon nanocylinders.

 The results reported here shed new light on the physics of nonradiating electromagnetic eigenmodes and provide a theoretical and experimental framework for designing accidental BICs\deleted{at will}. Our findings may enable the development of next-generation, tunable nanophotonic devices.

\begin{figure*}[htb!]
\includegraphics[width=1\textwidth]{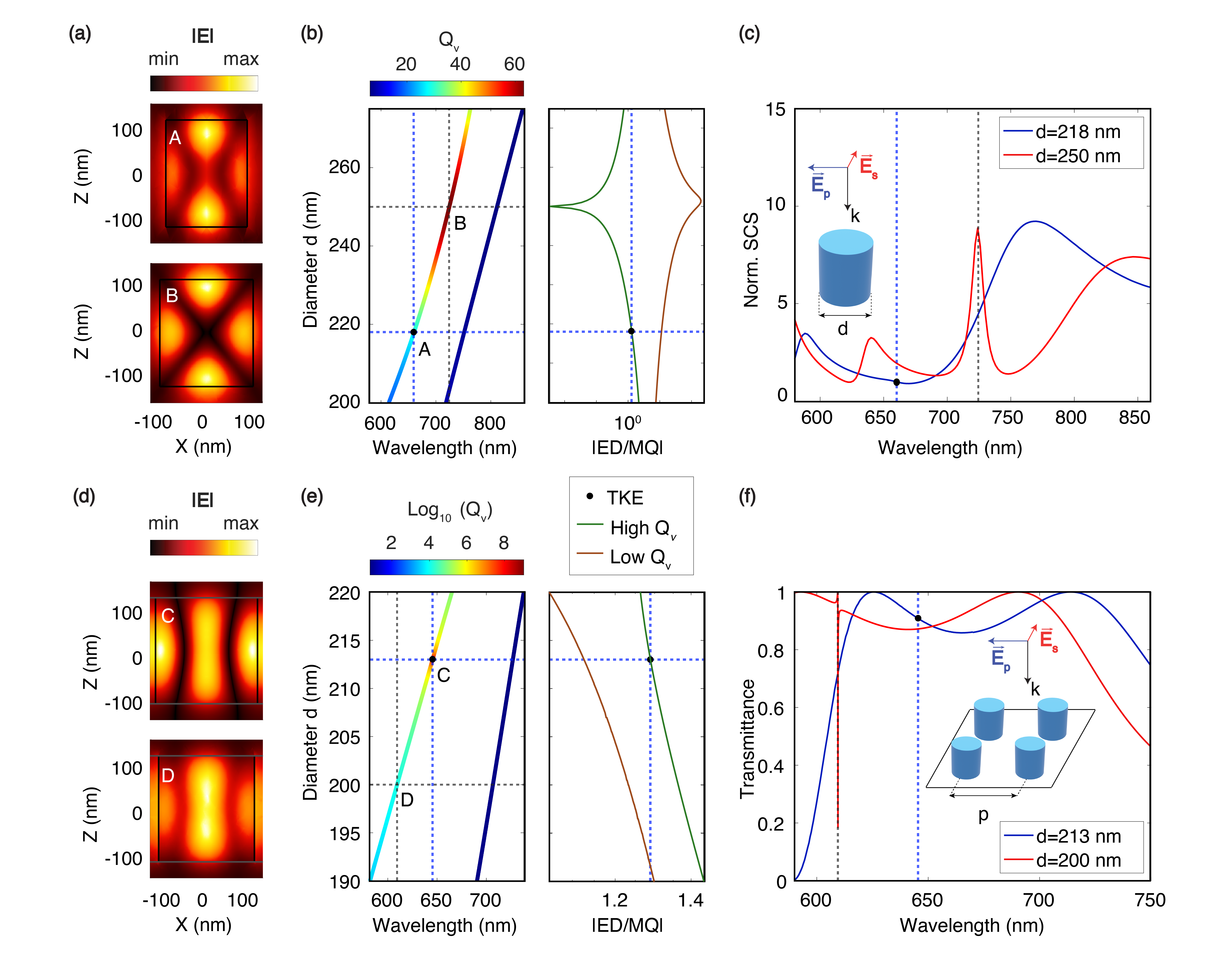}  
\caption{\added{Transverse Kerker effect (TKE) in a single dielectric cylinder and in a square array of dielectric cylinders. The nanoparticle height is  \replaced[id = SG]{$h = 230$}{$H = 230$} nm, the array period is \replaced[id = SG]{$p = 235.8$}{$P = 235.8$} nm, and the refractive index of the cylinder is $n = 3.85$. The surrounding material is air. \textbf{(a)} Electric field magnitude of eigenmodes in a single cylinder at the TKE (\replaced[id = SG]{A}{I}) and at the supercavity (\replaced[id = SG]{B}{II}) conditions. \textbf{(b)} Left: Dispersion of two eigenmodes (complex-valued resonance wavelength $\lambda_\nu$) with low and high quality factors $Q_\nu=|\text{Re}\lambda_\nu/2\text{Im}\lambda_\nu|$ (displayed by color). Right: $\mathrm{ED}/\mathrm{MQ}$ ratio of the multipole coefficients for these modes. \textbf{(c)} Normalized scattering cross section (SCS) of a single cylinder with \replaced[id = SG]{$d = 250$}{$D = 250$} nm (supercavity mode) and \replaced[id = SG]{$d = 218$}{$D = 218$} nm (TKE).
\textbf{(d)},\textbf{(e)} Same as in panel a but for an array of cylinders for the TKE condition (\replaced[id = SG]{C}{III}) and a quasi-BIC (\replaced[id = SG]{D}{IV}). \textbf{(f)} Spectra of the array: Transmittance for normally incident plane waves for \replaced[id = SG]{$d = 213$}{$D = 213$} nm (\replaced[id = SG]{C}{III}) and \replaced[id = SG]{$d = 200$}{$D = 200$} nm (\replaced[id = SG]{D}{IV}).} }
\label{fig:particleandarray}
\end{figure*}

\emph{Design of an ideal TKE source.}~We investigate the TKE beginning with a \emph{single} high-index particle and then proceed to periodic lattices, showing how symmetry and multipolar content of eigenmodes govern the suppression of out-of-plane radiation and the emergence of \replaced[id = TW]{states with high Q factors}{high-$Q$ states}. In the following, we use the \replaced[id=SG]{formalism}{language} of resonant states,~\cite{lalanne2018light, Kristensen2020a, both2021resonant} \added[id = TW]{also known as quasi-normal modes,} to describe the optical response of our structures. Resonant states are the natural eigenmodes of open photonic systems: \replaced[id = TW, comment = {Please ask yourself if resonant states can do something actively. For sure, they cannot solve Maxwell's equations ...}]{They are solutions to}{they solve} the macroscopic Maxwell equations in the absence of sources, subject to purely outgoing radiation boundary conditions. As a result of the latter, they have complex eigenfrequencies $\tilde{\omega}_\nu = \omega_\nu - i\gamma_\nu$, where $\omega_\nu$ and $\gamma_\nu > 0$ denote the resonance frequency and the \deleted[id = TW]{radiative} damping rate, respectively. \added[id = TW]{The damping rate consists of two contributions, which are radiative damping, i.e., loss due to photons that propagate away from the resonator, and intrinsic losses taking place in the resonator materials. Note that} $\omega_\nu$ and $\gamma_{\nu}$ are directly linked to the Q factor of the resonant state as $Q_{\nu} = \omega_{\nu}/2\gamma_{\nu}$. BICs correspond to the limiting case \replaced[id = TW]{of vanishing radiative loss. Therefore, they}{$\gamma_\nu \to 0$, i.e., resonant states that do not radiate, and therefore} can be considered to have an infinite Q factor in the absence of \replaced[id = TW]{material}{ohmic} losses.~\cite{canos2025exceptional} In the remainder of this work, we will refer to resonant states simply as ``eigenmodes.''

The eigenmodes of a system are solutions that exist regardless of \replaced[id = TW]{any}{the} illumination. Hence, there is no preferred ``forward'' direction. In the following, we focus only on those eigenmodes that can be excited by a normally incident plane wave. For this reason, it is important to fix what we call the forward radiation direction: We define it as the propagation direction of a normally incident plane wave impinging on the resonator from the top, \added[id = TW]{see Fig.~\ref{fig:particleandarray}}.

To delve into the radiation properties of each eigenmode, we perform a multipole expansion of their fields over vector spherical harmonics and derive the upward $P^{(+)}$ and downward $P^{(-)}$ radiated powers in terms of the electric and magnetic multipolar coefficients, following the guidelines in \added[id = TW, comment = {Are these refs. appropriate? Isn't the original paper from Carsten's group?}]{Refs.}~\citep{canos2023superscattering,kuznetsov2022special,wu2020intrinsic,alaee2019exact}. \replaced[id=SG]{The radiated powers can be written as}{The latter are found to be}
\begin{align}
P^{\text{(+)}} &\propto  \left| \sum_{l, m=\pm 1} (-i)^{l+1} \sqrt{2l+1} \left[ im\,a_{l,m} + b_{l,m} \right] \right|^2, \label{eq.upward_rp}\\
P^{\text{(-)}} &\propto \left|\sum_{l, m=\pm1}(i)^{l+1} \sqrt{2l+1} \left[ i m\,a_{l,m} -b_{l,m}\right] \right|^2 .\label{eq.upward_rm}
\end{align}
The coefficients $a_{l,m}$ and $b_{l,m}$ quantify the contributions of the electric and the magnetic multipole fields, respectively. The indices $l$ and $m$ are the angular momentum quantum numbers, with $l$ corresponding to the multipole order ($l=1$ as dipole, $l=2$ as quadrupole,~$\hdots$), and  $m$ \replaced[id = TW]{specifying the $z$ component of the total angular momentum. More specifically, left-handed circularly polarized plane waves that are incident from the top couple only to $m=1$ multipoles, while right-handed circularly polarized plane waves couple to $m=-1$}{the polarization of the multipole fields ($m=1$ is a left circularly polarized multipole and $m=-1$ is a right circularly polarized multipole)}. Additional details are provided in Section S1 of the Supporting Information. 

Importantly, we clarify that Eqs.~(\ref{eq.upward_rp}) and~(\ref{eq.upward_rm}) quantify only the intrinsic radiation by a single mode. Physically, they describe how much energy \emph{could be} radiated in the forward and backward directions if the mode field acted as a radiation source.~\cite{wu2020intrinsic} As a result, this expression \emph{is not limited by the optical theorem in any way}, unlike the total power scattered by the system upon plane wave excitation.

The TKE for a single mode would manifest if both the forward and backward radiation could be suppressed simultaneously. In most nanoresonators of interest in the optical range, the lowest order multipoles dominate, allowing the truncation of the expansions in Eqs.~(\ref{eq.upward_rp}) and (\ref{eq.upward_rm}) up to $l\leq 2$. 
For illustration, let us now consider an eigenmode whose far field properties are well described by the electric dipole (ED) term \(a_{1,\pm 1}\) and the magnetic quadrupole (MQ) term \(b_{2,\pm 1}\). Setting $P^{(+)}=P^{(-)}=0$ in Eqs.~(\ref{eq.upward_rp}) simultaneously  we derive a condition for an ideal TKE within the two-multipole approximation~\cite{Qin2022May}:

\begin{equation}\label{eq.TM}
\frac{\text{ED}}{\text{MQ}}=\frac{a_{1,1}-a_{1,-1}}{b_{2,1}+b_{2,-1}}=\sqrt{\frac{5}{3}}.
\end{equation}
An analogous condition can also be defined for eigenmodes having dominating contributions from the magnetic dipole and electric quadrupole terms [see Eq.~(S16) in Section S1 of the Supporting Information]. 
Equation~(\ref{eq.TM}) provides the phase and amplitude relations that the multipolar coefficients of the eigenmode of an isolated scatterer must fulfill in order to radiate light solely along the lateral directions. \added[id = TW]{Again,} we emphasize that this exact relation is impossible to achieve when regarding the light scattered by the object upon plane wave illumination, due to the optical theorem. However, the intrinsic radiation of the scatterer eigenmodes is not limited by it.

 To demonstrate that the eigenmodes of realistic systems can indeed support the ideal TKE, we now investigate a dielectric cylinder with refractive index close to that of silicon in the visible range, and vary its diameter \replaced[id = SG]{$d$}{$D$} [see Fig.~\ref{fig:particleandarray}\textbf{(a-c)}], allowing us to track the analytical dispersion of the eigenfrequencies and multipolar content of a few selected eigenmodes. Specifically, we select two eigenmodes whose radiation is primarily characterized by a mixture of ED and MQ contributions~\cite{koshelev2023resonant}, making them ideal candidates for achieving the TKE.

When varying the diameter of the cylinder, \deleted{it is well-known that }the hybridization between these two eigenmodes can give rise to a supercavity mode [see Fig.~\ref{fig:particleandarray}\textbf{(a)}].~\cite{Rybin2017Dec,Bogdanov2019Jan,Chen2019Sep,Koshelev2020Jan,Odit2021Jan} The latter corresponds to an eigenmode with strongly suppressed radiative losses, giving rise to a large Q factor (albeit not infinite). The increase in the Q factor is due to the suppression of radiation from the ED contribution, as can be seen in the right-most panel of Fig.~\ref{fig:particleandarray}\textbf{(c)}, where the ratio ED/MQ for \replaced[id = SG]{high $Q_\nu$ mode}{mode I} becomes close to zero for diameter values approaching the supercavity mode. As a result, the supercavity mode manifests in the scattering cross section as a narrow, symmetric Fano resonance [red dashed curve in  Fig.~\ref{fig:particleandarray}\textbf{(c)}]. 

Notably, we found that the hybridization of the two eigenmodes can also be exploited to precisely fulfill Eq.~(\ref{eq.TM}) for one of the eigenmodes at a given \replaced[id = SG]{$d$}{$D$}. It can be observed that the ratio ED/MQ of \replaced[id = SG]{high $Q_\nu$ mode}{mode I} grows linearly with \replaced[id = SG]{$d$}{$D$} far from the supercavity mode condition. This ratio reaches the desired value of \(\sqrt{5/3}\) at \replaced[id = SG]{$d=218$}{$D=218$}~nm for the mode with higher Q factor, (indicated by the black dot). The radiation from \replaced[id = SG]{high $Q_\nu$ mode}{mode I} then acquires a TKE nature, as confirmed by the calculated radiation pattern in Fig.~\ref{fig:particleandarray}\textbf{(b)}. As a result, the eigenmode can no longer couple to normally incident light and its resonant peak vanishes from the scattering spectra [blue line in Fig.~\ref{fig:particleandarray}\textbf{(c)}]. 

\emph{Emergence of TK BICs in metasurfaces.}~So far, we have shown that the eigenmodes of single nanoparticles can indeed exhibit a perfect TKE behavior, at the cost of no longer contributing to the scattering response under normally incident plane wave illumination. However, initially, the finding appears to be a pure theoretical curiosity. Next, we will show that, in fact, the TKE in a single eigenmode has \replaced[id = TW]{measurable}{observable} consequences in the optical observables, when the considered nanoparticles are used as building blocks for metasurfaces.

We now investigate a metasurface consisting of a square array of dielectric cylinders, [see Fig.~\ref{fig:particleandarray}\textbf{(d-f)}]. Despite the lattice interactions~\cite{rahimzadegan2022comprehensive}, the eigenmodes of the array are determined primarily by those of the single particle designed earlier. This can be visually appreciated upon comparing the eigenfield distributions of the single particle and the array in Fig.~\ref{fig:particleandarray}\textbf{(a)} and Fig.~\ref{fig:particleandarray}\textbf{(d)}.
As a result, \replaced[id = TW]{it is still possible to reach the TKE by tuning the diameter of the cylinder.}{tuning the diameter of the cylinder allows once more to reach the TKE.} However, in contrast to the single particle case, we observe that the Q~factor of the metasurface eigenmode diverges at the TKE condition, giving rise to an accidental BIC. The transmission coefficient as a function of wavelength reveals no resonance at the BIC spectral position [see Fig.~\ref{fig:particleandarray}\textbf{(f)}]. 

A natural question arises: Why does the TKE yield an infinite Q~factor for the array, and why does it not simply correspond to the maximum for a single particle?
A detailed answer to this question is provided in Section~S2 of the Supporting Information. We show that the Q factor of an eigenmode is inversely proportional to the radiated power in all directions $P_{\text{tot}}$, so that $Q\propto 1/P_{\text{tot}}$. Therefore, in the single particle case, despite radiation being suppressed in the forward and backward directions, light can still radiate in all the other directions, resulting in $P_{\text{tot}}\neq 0$ and a finite Q~factor. In contrast, the eigenmodes in a metasurface can only radiate away in the forward and backward directions, so that $P_{\text{tot}} = P^{(+)} + P^{(-)}$. Thus, at the TKE $P_{\text{tot}} = 0$ by definition, and the Q~factor diverges \added[id = TW]{in the absence of material losses. In the Supporting Information (see Section S3), we demonstrate by the perturbative approach in Ref.}~\citep{Yan2020Jul} \added{that these states arise by the hybridization of resonant states, as in the single particle case.}\deleted{As demonstrated in the Supporting Information (see Section S3) by using the resonant state perturbative approach in Ref.~\cite{Yan2020Jul}, these high-Q states also arise from the hybridization of resonant states, as in the single particle case.}
 
\replaced[id=SG]{While the present discussion primarily investigates}{While we have investigated} eigenmodes where the electric dipole and magnetic quadrupole are dominant, we emphasize that other multipolar combinations can also give rise to the TKE. For instance, Fig.~\ref{fig:concept}\textbf{(c)} shows cases in which the magnetic dipole and the electric quadrupole play the leading role.  Moreover, the TKE can also be realized in higher-order resonances, as demonstrated in the Supporting Information (see Section~S4). We further performed a symmetry-based classification of the eigenmodes that can host these TK BICs, which clarifies the underlying selection rules and modal degeneracies (see Section~S5).

\begin{figure*}
\includegraphics[width=1\textwidth]{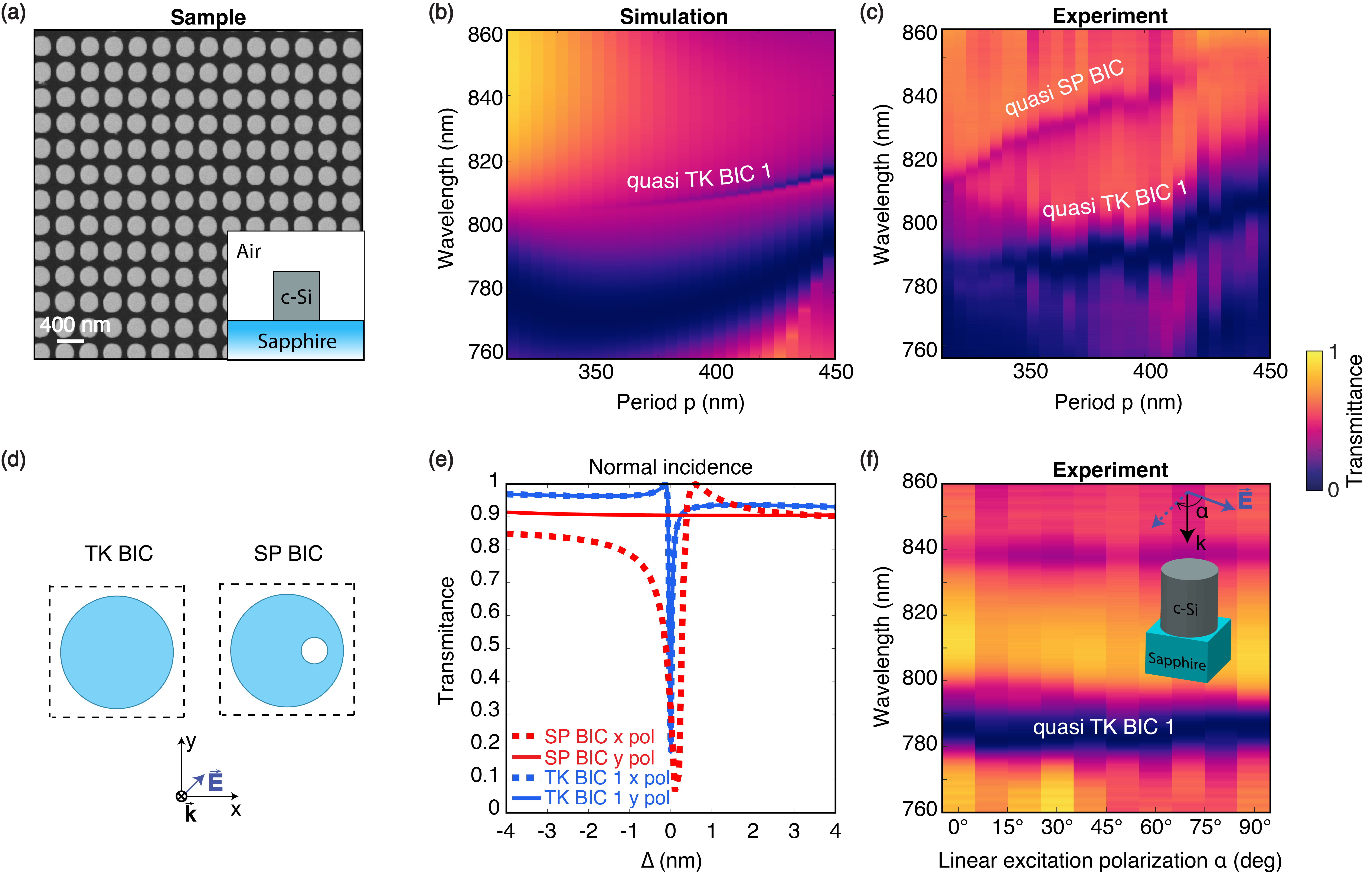}
\caption{Observation of the transverse Kerker bound state in the continuum (TK BIC) at optical wavelengths and its polarization robustness. 
\textbf{(a)} SEM image of the fabricated silicon nanodisk metasurface on a sapphire substrate (disk diameter \replaced[id = SG]{$d=275~\mathrm{nm}$}{$D=275~\mathrm{nm}$}, height \replaced[id = SG]{$h=300~\mathrm{nm}$}{$H=300~\mathrm{nm}$}).
\textbf{(b)} Simulated and \textbf{(c)} measured normal incidence transmittance spectra versus wavelength and lattice period. 
\textbf{(d)} Unit-cell geometries\added[id=SG]{, adapted from Fig.~2,} supporting two different BIC mechanisms: TK BIC (left)—a square array of dielectric cylinders with refractive index $n=3.85$—and a \added{quasi-}BIC (right) realized by introducing a symmetry-breaking defect, i.e., a hole aligned along the $y$ axis with radius $r=10~\mathrm{nm}$. Cylinder diameter \replaced[id = SG]{$d=200~\mathrm{nm}$}{$D=200~\mathrm{nm}$}, height \replaced[id = SG]{$h=230~\mathrm{nm}$}{$H=230~\mathrm{nm}$}, period \replaced[id = SG]{$p=235.8~\mathrm{nm}$}{$P=235.8~\mathrm{nm}$}, refractive index $n= 3.85$.
\textbf{(e)} Simulated normal incidence transmission spectra plotted versus wavelength detuning from the resonance for $x$- and $y$-polarized illumination: the TK BIC resonance position is identical for both polarizations, whereas the SP BIC resonance is present only for $y$ polarization. 
\textbf{(f)} Experimental transmission map as a function of linear-polarization angle $\alpha$ and wavelength for a metasurface of silicon cylinders on sapphire at fixed period \replaced[id = SG]{$p=400~\mathrm{nm}$}{$P=400~\mathrm{nm}$}, confirming that the quasi TK BIC resonance is preserved under rotation of the incident polarization.}
\label{fig:experiment}
\end{figure*}

\emph{Experimental demonstration.}~To validate our theoretical predictions, we have fabricated a metasurface of silicon nanodisks on a sapphire substrate and performed optical transmission measurements. The scanning electron microscopy (SEM) image of the fabricated sample is shown in Fig.~\ref{fig:experiment}\textbf{(a)}. A detailed description of the fabrication process and measurement setups is provided in Sections S6 and S7 of the Supporting Information. The experimental transmittance spectra as a function of wavelength and lattice period under normal incidence excitation are presented in Fig.~\ref{fig:experiment}\textbf{(c)}, while the corresponding numerical results obtained \replaced[id = TW]{via the Fourier modal method~\cite{Weiss2009a} (also known as rigorous coupled-wave analysis)}{using the S-matrix method (RCWA)} are displayed in Fig.~\ref{fig:experiment}\textbf{(b)}. The measurements reproduce the simulated trends and reveal polarization-independent \replaced[id = TW]{resonances with high Q factors}{high-Q resonance} in the range of 780–810~nm, exhibiting a blueshift of approximately 20 nm compared to the simulations, likely due to fabrication \replaced[id=SG]{inaccuracies.}{tolerances.} An additional resonance feature is observed in the range of 810–860 nm, which corresponds to a symmetry-protected (SP) BIC mode. The underlying nature of this resonance is a vertical electric dipole that ideally cannot be excited under normal incidence illumination. However, due to the \cor{finite numerical aperture of the incident beam} in the experiment, this mode becomes weakly accessible and is therefore visible in the transmission map (see \added[id = TW]{numerical validation in} Section S8 \replaced[id = TW]{of}{in} the Supporting Information). 

\emph{Polarization-independent BIC.} As discussed above, \replaced[id=SG]{the quasi-BIC emerging from the TK BIC}{the TK BIC} is polarization independent. Figure~\ref{fig:experiment}\textbf{(d-f)} illustrates this property clearly. In Fig.~\ref{fig:experiment}\textbf{(d)}, two unit-cell designs are shown for two different BIC types: the TK BIC and the symmetry-protected BIC. The left design corresponds to the previously considered structure, a square metasurface composed of high-index dielectric cylinders, whereas the right design introduces a symmetry-breaking defect in the form of a narrow hole oriented along the $y$~axis, which enables a polarization-selective symmetry-protected quasi-BIC. \added[id=SG]{In contrast, to obtain a quasi-BIC emerging from the TK BIC, it is sufficient to detune the geometric parameters of the system, which shifts the mode away from the ideal BIC condition and results in a resonance with finite radiative Q factor.}

Figure~\ref{fig:experiment}\textbf{(e)} presents the normal incidence transmission spectra plotted versus the wavelength detuning centered at the resonance. For the TK BIC, the resonance position coincides for $x$- and $y$-polarized incident waves, confirming that the response does not depend on the linear polarization at normal incidence. In contrast, the symmetry-protected BIC-related resonance appears only for the $y$~polarization, evidencing strong polarization selectivity induced by the broken symmetry. \deleted[id = TW]{Finally,} Figure~\ref{fig:experiment}\textbf{(f)} shows an experimental transmission map as a function of polarization angle $\alpha$ and wavelength for the same metasurface platform (silicon cylinders on a sapphire substrate), demonstrating that the TK BIC resonance indeed remains essentially unchanged when the incident linear polarization is rotated, which is consistent with a polarization-independent response. \cor{We emphasize that, in order to obtain polarization-independent \added{quasi-}BICs, prior works relied on complex metasurface designs enabled by Brillouin-zone folding~\cite{yang2025polarization,vaity2022polarization}.}

\begin{figure*}
\includegraphics[width=1\textwidth]{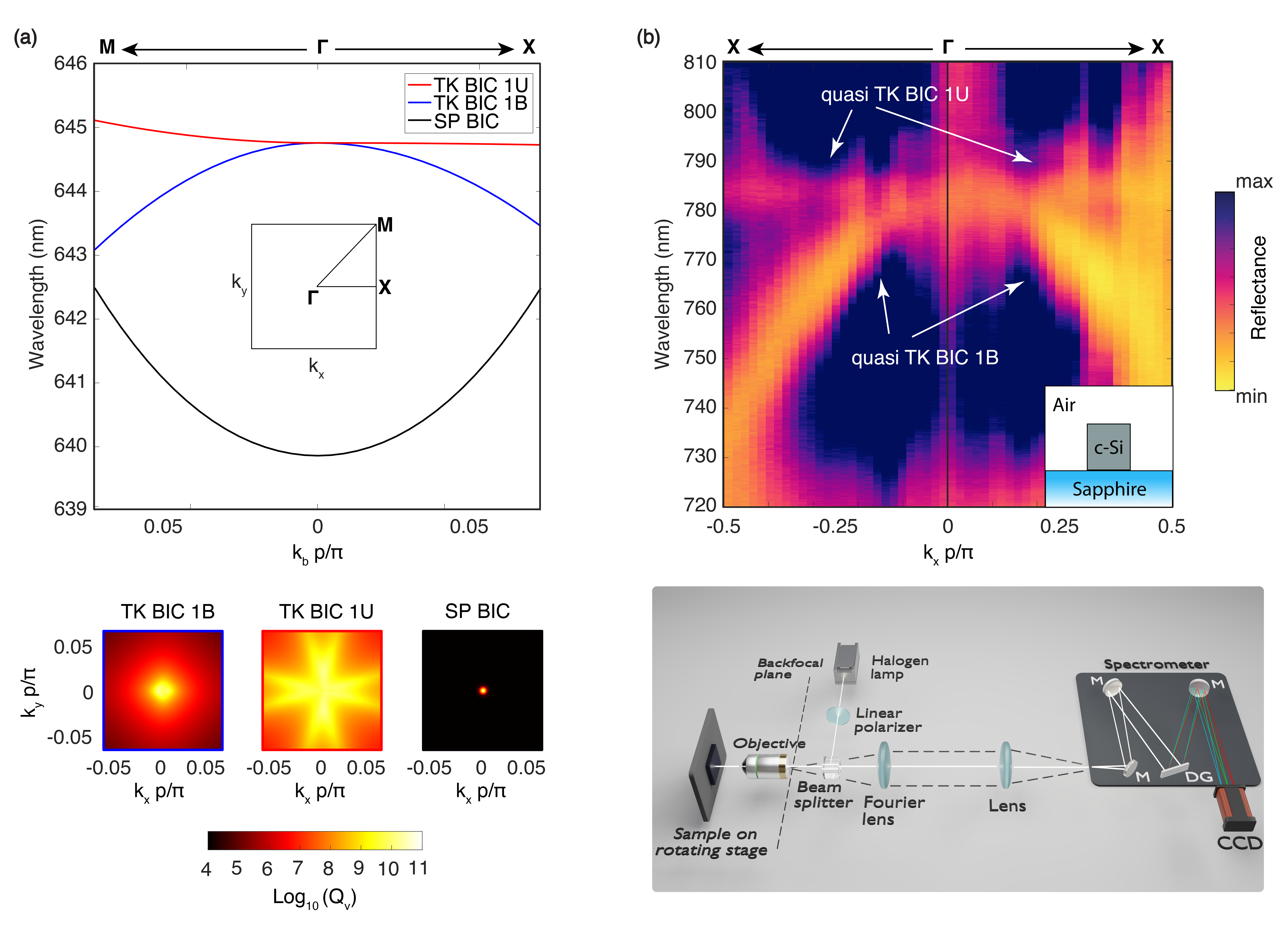}
\caption{\replaced[id=SG]{TK BICs in momentum Space}{Momentum-space robustness of TK BICs near the $\Gamma$ point}.
\textbf{(a)} Numerical analysis of the stability of Transverse-Kerker bound states in the continuum (TK BICs) around the $\Gamma$ point, benchmarked against a conventional symmetry-protected (SP) BIC supported by a vertical electric-dipole-like mode. Dispersion (resonance wavelength versus in-plane Bloch wavevector) in the vicinity of $\Gamma$ along the $\Gamma$X and $\Gamma$M directions. \replaced[id=SG]{Q factor}{Q-factor} maps around $\Gamma$, showing that the two degenerate TK BIC modes (TK BIC \replaced[
id=SG]{1U}{1} and TK BIC \replaced[
id=SG]{1B}{2}) exhibit a substantially slower radiative-Q degradation with increasing $|{\bf k}_\text{b}|$ than the SP BIC. Simulated structure: square array of dielectric cylinders with refractive index $n=3.85$, height \replaced[id=SG]{$h=230~\mathrm{nm}$}{$H=230~\mathrm{nm}$}, period \replaced[id=SG]{$p=235.8~\mathrm{nm}$}{$P=235.8~\mathrm{nm}$}, and diameter \replaced[id=SG]{$d=213~\mathrm{nm}$}{$D=213~\mathrm{nm}$}.
\textbf{(b)} Experimental angle-resolved reflection of a silicon-nanodisk metasurface on a sapphire substrate, presented as a reflection map  \replaced[id=SG]{scanned along $X-\Gamma-X$ (varying the in-plane wavevector $k_x$ through $\Gamma$ at $k_y=0$)}{versus Bloch wavevector $k_\text{b}$ scanned along $\Gamma$X and $-X$ $\Gamma$}. Two dispersive branches are resolved, corresponding to the two modes that are degenerate at $\Gamma$ and constitute the quasi-BIC at normal incidence. 
Schematic of the angle-resolved reflection setup: a halogen lamp provides collimated white light, linearly polarized by a linear polarizer and focused onto the sample through a $60\times$ (NA$=0.95$) objective via a $50:50$ beamsplitter. The reflected beam is collected by the same objective; a Fourier lens images the backfocal plane, and a 4f relay projects it onto the spectrometer entrance slit. The dispersed signal is recorded on a CCD (M, mirror).}

\label{fig:robustness}
\end{figure*}

\emph{Robustness of TK BICs near the $\Gamma$ point.}
Figure~\ref{fig:robustness} compares the momentum-space robustness of \deleted[id = TW]{Transverse Kerker BICs (}TK BICs near the $\Gamma$ point with a symmetry-protected (SP) BIC associated with a vertically oriented electric dipole. In Fig.~\ref{fig:robustness}\textbf{(a)}, we first plot the dispersion in the vicinity of $\Gamma$, along the $\Gamma$X and $\Gamma$M directions. The corresponding Q~factor maps in $k$-space reveal that the two degenerate modes (TK BIC \replaced[
id=SG]{1U}{1} and TK BIC \replaced[
id=SG]{1B}{2}) retain high radiative $Q$ over a substantially broader region around $\Gamma$ than the mode with SP BIC. \cor{This enhanced robustness can be interpreted as the result of a merging process. Specifically, we show in Sections~S9-S10 of the Supporting Information how multiple off-$\Gamma$ accidental BICs merge at the $\Gamma$ point, once the Transverse Kerker condition is fulfilled.} Away from the $\Gamma$ point, the degeneracy between TK BIC \replaced[
id=SG]{1U}{1} and TK BIC \replaced[
id=SG]{1B}{2} is lifted; consequently, TK BIC \replaced[
id=SG]{1B}{1} exhibits a faster \replaced[id = TW]{decay of the Q~factor}{$Q$ decay} than \replaced[
id=SG]{1U}{2}, while still maintaining very high \replaced[id = TW]{Q factors}{$Q$ values}.  Hence, we conclude that TK BICs correspond to ``super-BICs,'' which have been shown to lower the lasing threshold in finite-sized metasurfaces.~\cite{hwang2021ultralow}. The topological characterization supporting these conclusions, as well as a detailed analysis of the BIC-merging process, is provided in the Supporting Information (see Sections S9-S10).
In Fig.~\ref{fig:robustness}\textbf{(b)}, we experimentally probe the same physical picture using the fabricated silicon cylinders on sapphire. The measured reflection map is shown as a function of the Bloch wavevector ${\bf k}_\text{b}$ scanned along \replaced[id=SG]{$X-\Gamma-X$}{$-X$ $\Gamma$ and along $\Gamma$ $X$}. Two clearly resolved dispersive branches are observed, corresponding to the two modes that become degenerate at $\Gamma$ and form a quasi-BIC at normal incidence. The schematic below the map depicts the angular-resolved reflection measurement configuration. \deleted{used to access the near-$\Gamma$ dispersion.}

In summary, we establish the transverse Kerker effect (TKE) as a practical and predictive design principle for creating accidental bound states in the continuum (BICs) in simple, fully symmetric dielectric metasurfaces.  By linking the intrinsic multipolar composition of a single eigenmode to the simultaneous suppression of upward and downward radiation, we identify a route to an ideal TKE and show that, once embedded into a periodic lattice, this condition eliminates the available out-of-plane channels at the $\Gamma$ point and enforces a transverse-Kerker BIC under normal incidence.  In contrast to symmetry-protected BICs, this mechanism yields a doubly degenerate and therefore polarization-independent quasi-BIC that can be accessed without symmetry breaking or Brillouin-zone folding, while retaining straightforward unit-cell geometry and broad tunability through geometric detuning. These features directly address two persistent practical limitations of high-$Q$ metasurface resonances: Polarization selectivity and stringent constraints on alignment and numerical aperture.
Beyond the existence of \deleted[id = TW]{a high-$Q$} resonances \added[id = TW]{with high Q~factors} at $\Gamma$, we further show that transverse-Kerker BICs maintain elevated Q~factors over a substantially wider region of momentum space than symmetry-protected BICs, implying improved tolerance to finite-NA excitation and moderate wavevector mismatch. Experimentally, we validate the concept in the visible using silicon nanodisk metasurfaces on sapphire, observing a polarization-independent high-$Q$ feature consistent with the theory, \cor{and confirm their super-BIC character through momentum-resolved optical measurements}. Looking forward, transverse Kerker BICs provide a promising platform for low-threshold lasers, flat active devices and enhanced nonlinear and chiroptical interactions.

\begin{acknowledgement}
Thomas Weiss and Sergei Gladyshev acknowledge funding by the individual DFG project WE5815/5-1. 
Connor Heimig, Tao Jiang, Angana Bhattacharya and Andreas Tittl acknowledge funding by the European Union (EIC, NEHO, 101046329, ERC, METANEXT, 101078018). Views and opinions expressed are however those of the author(s) only and do not necessarily reflect those of the European Union or the European Research Council Executive Agency. Neither the European Union nor the granting authority can be held responsible for them. This project was also funded by the Deutsche Forschungsgemeinschaft (DFG, German Research Foundation) under grant numbers EXC 2089/1-390776260 (Germany’s Excellence Strategy) and TI 1063/1 (Emmy Noether Program), the Bavarian program Solar Energies Go Hybrid (SolTech) and the Center for NanoScience (CeNS).
\end{acknowledgement}

 \begin{suppinfo}

The Supporting Information provides the theoretical and numerical details underlying the transverse-Kerker BIC concept, including the multipolar expansion of eigenmodes and closed-form expressions linking the mode’s multipole content to upward/downward radiation, from which the ideal TKE conditions are derived (Sec.~S1). It further clarifies the relation between radiative losses, the imaginary part of the eigenfrequency, and why the Q-factor can diverge only in a periodic array at the $\Gamma$ point (Sec.~S2). Additional modeling is presented via resonant-state expansion for the formation of the accidental TK BIC (Sec.~S3), together with a full normal incidence transmittance map revealing higher-order TK-BIC families and practical design windows (Sec.~S4). 
The Supporting Information also contains a symmetry classification of TKE-type eigenmodes across lattice point groups (Sec.~S5). Fabrication and optical-measurement workflows and supplementary transmission spectra (including angular averaging) are reported to support the experimental observations (Secs.~S6--S8). Finally, the Stokes-parameter formalism, polarization maps, and the topological charges of the associated polarization vortices are presented in Secs.~S9--S10.

 \end{suppinfo}

\bibliography{TKE}

\end{document}